\documentclass[a4paper]{article}

\usepackage{INTERSPEECH2022}
\usepackage{multirow}

\title{Exploring the influence of fine-tuning data on wav2vec 2.0 model for blind speech quality prediction}
\name{Helard Becerra, Alessandro Ragano, Andrew Hines}
\address{
  School of Computer Science, University of College Dublin, Dublin, Ireland}
\email{helard.becerra@ucd.ie, alessandro.ragano@ucdconnect.ie, andrew.hines@ucd.ie}

\begin{document}

\maketitle
\begin{abstract}
Recent studies have shown how self-supervised models can produce accurate speech quality predictions. Speech representations generated by the pre-trained wav2vec 2.0 model allows constructing robust predicting models using small amounts of annotated data. This opens the possibility of developing strong models in scenarios where labelled data is scarce. It is known that fine-tuning improves the model's performance; however, it is unclear how the data (e.g., language, amount of samples) used for fine-tuning is influencing that performance. In this paper, we explore how using different speech corpus to fine-tune the wav2vec 2.0 can influence its performance. We took four speech datasets containing degradations found in common conferencing applications and fine-tuned wav2vec 2.0 targeting different languages and data size scenarios. The fine-tuned models were tested across all four conferencing datasets plus an additional dataset containing synthetic speech and they were compared against three external baseline models. Results showed that fine-tuned models were able to compete with baseline models. Larger fine-tune data guarantee better performance; meanwhile, diversity in language helped the models deal with specific languages. Further research is needed to evaluate other wav2vec 2.0 models pre-trained with multi-lingual datasets and to develop prediction models that are more resilient to language diversity.
\end{abstract}

\section{Introduction}

Speech quality assessment is key to monitoring and evaluating the performance of applications and services in which speech is an essential component. Their success depends heavily on assuring acceptable levels of perceived quality, thus the importance of measuring it. Accurate quality assessments can be gathered through subjective experiments in the form of mean opinion scores (MOS) \cite{rec1996p} or using the MUltiple Stimuli with Hidden Reference and Anchor method (MUSHRA) \cite{recommendation2001method}, yet, the resources required (e.g., time, human availability) and their nature make it hard to implement them over an automatic quality monitoring design. Computer algorithms (objective metrics) that can automatically estimate speech quality offer a practical solution to this problem. Within speech objective metrics, Full-reference (FR) metrics, which use both degraded and reference speech signals, have been shown to work very well at estimating the perceived speech quality \cite{hines2012speech, beerends2013perceptual}; however, they depend on the availability of the reference signal. On the other hand, No-reference metrics (NR), which rely solely on the degraded signal, can be deployed in real-time monitoring scenarios, but at the cost of accuracy \cite{malfait2006p}.

Machine learning (ML) provides the potential for \mbox{data-driven} metrics rather than similarity or distance error-based approaches that have been shown to work well in FR scenarios \cite{hines2012speech}. Self-supervised pre-trained speech models have shown to be very useful for different speech-related tasks, including speech quality prediction \cite{tseng2021utilizing, cooper2021generalization}. The wav2vec 2.0 model, a self-supervised learning (SSL) framework composed of a multi-layer convolutional feature encoder and a Transformer, was successfully applied for MOS prediction using the fixed-size representations generated by the model \cite{cooper2021generalization, tseng2021utilizing}. Models based on deep-learning techniques have also been developed to predict the speech quality with very accurate outcomes. NISQA is a model based on a convolutional neural network (CNN) for frame-wise feature computation, a self-attention block for time-dependency modelling, and an attention-pooling for estimating the overall MOS. This model showed stable results when tested on unknown data and live phone calls \cite{mittag2021nisqa}. Despite their promising results, data-driven models are exposed to bias depending on the type of data used to train them. Collecting data that does not bias the model is a challenge (specifically neutral to: speaker voice/accent, language, degradation) \cite{jimenez2021removing,Jimenez21language,chinen21marginal}.

Several studies have pointed out the importance of the fine-tuning step to achieve better performance, especially over specific contexts where labelled data is scarce or hard to collect \cite{fan2020exploring, cooper2021generalization}. For wav2vec 2.0, adding a fine-tune step has shown a noticeable accuracy improvement in MOS prediction over different speech datasets \cite{tseng2021utilizing, cooper2021generalization}. Adding this step facilitates the exploration of different testing scenarios involving specific types of degradation, speech languages, or subjective collection methods (e.g., MUSHRA, MOS). However, further research is needed to understand the level of influence that the data (e.g., language, type of distortion, amount of samples) used for fine-tuning is exerting over the model's performance.

This paper aims: 1) to better understand the role of fine-tuning in wav2vec 2.0 based models and how the content and size of the fine-tuning datasets impact the model performance over a range of speech datasets, 2) to compare wav2vec 2.0 based models and NISQA based models over a range of speech datasets.
With that aim, two experiments were carried out using the fine-tuning system presented in \cite{cooper2021generalization}. The first experiment explores the content language aspect; for this experiment, three models were fine-tuned using only English, German, and Chinese speech samples. The experiment aims to better understand the influence of language on speech quality prediction on datasets with different language samples. The second experiment explores the size of the datasets used for fine-tuning; three models were fine-tuned using 1 k, 10 k, and 50 k mixed speech samples. The experiment seeks to understand the level of influence that fine-tuning data size has on speech quality predictions. Labelled speech datasets containing conferencing degradations (Tencent, IU Bloomington \cite{stupakov2009cosine,richey2018voices}, NISQA Corpus \cite{mittag2021nisqa}, and PSTN \cite{mittag2020dnn}) and synthesised speech (VoiceMOS \cite{cooper2021generalization}) were used to build the fine-tuning datasets and test the resulting models.

\section{Methods}

\subsection{The wav2vec 2.0 pre-trained model}
This model is based on the wav2vec 2.0 SSL framework, which consists of a convolutional feature encoder and a Transformer that takes raw waveforms as input and produces fixed-size speech representations \cite{baevski2020wav2vec}. The framework was successfully applied to speech-related tasks like speech recognition and language verification \cite{fan2020exploring, baevski2020effectiveness}. Due to a large amount of data on which the model is pre-trained, only a small amount of data is required to apply the model over different tasks on which speech representation plays a key role. The promising results presented in prior studies, more specifically for speech quality prediction \cite{cooper2021generalization,tseng2021utilizing}, plus the release of annotated datasets for speech quality challenges, motivated the exploration of this framework and how fine-tuning can influence its performance in the speech quality prediction task. The wav2vec 2.0 model, pre-trained with 53.2 k hours of audio from the Librispeech dataset \cite{panayotov2015librispeech}, served as the base to fine-tune a set of models targeting specific speech contexts as presented in Table~\ref{tab:experiments}.

\subsection{Datasets}

The availability of large multi-lingual labelled datasets, in the context of the VoiceMOS and ConferencingSpeech challenges, enabled the realization of this study. Such datasets include more than 200 hours of speech samples degraded with common degradations experienced over conferencing applications plus speech synthesis and voice conversion samples. This study considered the synthetised speech dataset VoiceMOS \cite{cooper2021generalization}, plus four sets of datasets with speech conferencing distortions, namely: Tencent (2 datasets), NISQA (7 datasets) \cite{mittag2021nisqa}, IU-Bloomington (2 datasets)\cite{stupakov2009cosine,richey2018voices} and PSTN \cite{mittag2020dnn}. These datasets were used to build subsets to fine-tune the wav2vec 2.0  pre-trained model targeting specific speech scenarios. Table \ref{tab:datasets} presents the main characteristics of the 13 datasets used in this study.

\subsection{Fine-tune experiments}

Two experiments were conducted to help understand the impact of fine-tuning data over the wav2vec 2.0 performance: fine-tuning models based on the datasets' language and size. The language models were fine-tuned using samples containing speech in German, Chinese, and English. This experiment aims to understand how the language applied during the fine-tuning is affecting the models’ performance. For the second group of models, fine-tuning was done using different amounts of speech samples: 1 k, 10 k and 50 k. This experiment aims to observe how different datasets sizes used for fine-tuning the models influence their performance. Train and test subsets were randomly generated using an 80-20 percent ratio and maintaining a similar MOS distribution across all fine-tune datasets. The wav2vec 2.0 pre-trained model~\footnote{https://github.com/pytorch/fairseq/tree/main/examples/wav2vec} and the fine-tune SSL system released for the VoiceMOS challenge~\footnote{https://github.com/nii-yamagishilab/mos-finetune-ssl} were used to fine-tune the models for the two experiments. In this system, fine-tuning is done by adding an output mean-pooling layer trained with an L1 loss function. The mean square error (MSE) between the predicted and target MOS is monitored on each training epoch, and training is stopped if no improvement is observed after 20 epochs. Details about the fine-tuned models and the subsets used to build them are presented in Table \ref{tab:experiments}.

% Table generated by Excel2LaTeX from sheet 'experiments'
\begin{table}[t]
  \centering
  \caption{Fine-tune description for Experiments 1 and 2. all(L): German, Chinese and English. all(D): NISQA TestLiveTalk, Tencent, and PSTN. PT: pre-trained. FT: fine-tuned.}
    \resizebox{1\columnwidth}{!}{
    \begin{tabular}{l|ll|lrrr}
    \hline
          & \multicolumn{2}{c|}{\textbf{Models}} & \multicolumn{4}{c}{\textbf{Fine-tune Datasets}} \\
    \textbf{Experiment} & \textbf{FT Model} & \textbf{PT Model} & \textbf{Language} & \multicolumn{1}{l}{\textbf{Train samples}} & \multicolumn{1}{l}{\textbf{Val samples}} & \multicolumn{1}{l}{\textbf{Source Dataset}} \\
    \hline
    Exp. 1 & w2v\_ger & wav2vec 2.0 & German & 185   & 47    & \multicolumn{1}{l}{Nisqa TestLiveTalk} \\
    Exp. 1 & w2v\_cn & wav2vec 2.0 & Chinese & 2424  & 606   & \multicolumn{1}{l}{Tencent} \\
    Exp. 1 & w2v\_en & wav2vec 2.0 & English & 7515  & 1879  & \multicolumn{1}{l}{PSTN} \\
    Exp. 1 & w2v\_all & wav2vec 2.0 & all(L)   & 11429 & 2858  & \multicolumn{1}{l}{all(D)} \\
    \hline
    Exp. 2 & w2v\_1k & wav2vec 2.0 & English & 1884  & 472   & \multicolumn{1}{l}{PSTN/Tencent} \\
    Exp. 2 & w2v\_10k & wav2vec 2.0 & English & 8995  & 2248  & \multicolumn{1}{l}{PSTN/Tencent} \\
    Exp. 2 & w2v\_50k & wav2vec 2.0 & English & 56217 & 14005 & \multicolumn{1}{l}{PSTN/Tencent} \\
    \hline
    Baseline & w2v\_base & wav2vec 2.0 & English & 4974  & 1066  & \multicolumn{1}{l}{VoiceMOS} \\
    \hline
    Baseline & nq\_base1 &  & English & 70764 & 9713 &  Nisqa/Tencent/PSTN \\
    Baseline & nq\_base2 &  & English & 70764 & 9713 &  Nisqa/Tencent/PSTN \\
    \hline
    \end{tabular}}%
  \label{tab:experiments}%
\end{table}%

% Table generated by Excel2LaTeX from sheet 'datasets'
\begin{table*}[htbp]
  \centering
  \caption{Description of the speech datasets.}
    \resizebox{2\columnwidth}{!}{
    \begin{tabular}{llllrrrcc}
    \hline
    \textbf{Dataset} & \textbf{ID} & \textbf{Language} & \textbf{Degradation Types} & \multicolumn{1}{l}{\textbf{Samples}} & \multicolumn{1}{l}{\textbf{Speakers}} & \multicolumn{1}{l}{\textbf{Systems}} & \multicolumn{1}{l}{\textbf{Methodology}} & \multicolumn{1}{l}{\textbf{Ratings per sample}} \\
    \hline
    VoiceMOS & VoMOS & English & test-to-speech, voice conversion, natural speech & 7106  & 27    & 187   & MOS   & 8 \\
    IU-Bloomington Cosine & IUCo  & English & Background noise & 18000 &       &       & MUSHRA &  \\
    IU-Bloomington Voices & IUVo  & English & Reverberation & 18000 &       &       & MUSHRA &  \\
    \multirow{2}[0]{*}{Tencent w/o Reverberation} & \multirow{2}[0]{*}{TcwoR} & \multirow{2}[0]{*}{Chinese} & \multicolumn{1}{p{27.215em}}{white noise, background noise, } & \multirow{2}[0]{*}{8366} & \multirow{2}[0]{*}{} & \multirow{2}[0]{*}{} & \multirow{2}[0]{*}{MOS} & \multirow{2}[0]{*}{} \\
          &       &       & frequency filtering, amplitude clipping, codecs &       &       &       &       &  \\
    Tencent w Reverberation & TCwR  & Chinese & Reverberation & 3197  &       &       & MOS   & ~24 \\
    \multirow{2}[0]{*}{NISQA Corpus TestFor} & \multirow{2}[0]{*}{NTsF} & \multirow{2}[0]{*}{English} & Codecs, background noises, packet-loss, clipping, & \multirow{2}[0]{*}{240} & \multirow{2}[0]{*}{80} & \multirow{2}[0]{*}{60} & \multirow{2}[0]{*}{MOS} & \multirow{2}[0]{*}{~30} \\
          &       &       & live conditions with WhatsApp, Zoom, and Discord &       &       &       &       &  \\
    NISQA Corpus TestLiveTalk & NTsLT & German & Natural environment conditions & 232   & 8     & 58    & MOS   & 24 \\
    \multirow{3}[0]{*}{NISQA Corpus TestP501} & \multirow{3}[0]{*}{NTsP501} & \multirow{3}[0]{*}{English} & Codecs, background noises, packet-loss, clipping, & \multirow{3}[0]{*}{240} & \multirow{3}[0]{*}{4} & \multirow{3}[0]{*}{60} & \multirow{3}[0]{*}{MOS} & \multirow{3}[0]{*}{~28} \\
          &       &       & live conditions with Skype, Zoom, WhatsApp, &       &       &       &       &  \\
          &       &       & and mobile network recordings &       &       &       &       &  \\
    NISQA Corpus TrainLive & NTrL  & English & live telephone and Skype calls & 1020  & 486   & 1020  & MOS   & ~5 \\
    \multirow{3}[0]{*}{NISQA Corpus TrainSim} & \multirow{3}[0]{*}{NTrS} & \multirow{3}[0]{*}{English} & white gaussian noise, MNRU noise, noise clips, & \multirow{3}[0]{*}{10000} & \multirow{3}[0]{*}{2322} & \multirow{3}[0]{*}{10000} & \multirow{3}[0]{*}{MOS} & \multirow{3}[0]{*}{~5} \\
          &       &       & frequency filters, amplitude clipping, speech level changes, &       &       &       &       &  \\
          &       &       & codecs, packet-loss &       &       &       &       &  \\
    NISQA Corpus ValLive & NVaL  & English & live telephone and Skype calls & 200   & 102   & 200   & MOS   & ~5 \\
    \multirow{3}[0]{*}{NISQA Corpus ValSim} & \multirow{3}[0]{*}{NVaS} & \multirow{3}[0]{*}{English} & white gaussian noise, MNRU noise, noise clips, & \multirow{3}[0]{*}{2500} & \multirow{3}[0]{*}{2500} & \multirow{3}[0]{*}{2500} & \multirow{3}[0]{*}{MOS} & \multirow{3}[0]{*}{~5} \\
          &       &       & frequency filters, amplitude clipping, speech level changes, &       &       &       &       &  \\
          &       &       & codecs, packet-loss &       &       &       &       &  \\
    PSTN  & PSTN  & English & Background noise & 58709 & 2150  &       & MOS   & 5 \\
    \hline
    \end{tabular}}%
  \label{tab:datasets}%
\end{table*}%

\section{Results}

\begin{figure*}[bt]
\centering
\begin{tabular}{ccc}
\includegraphics[width=0.65\columnwidth]{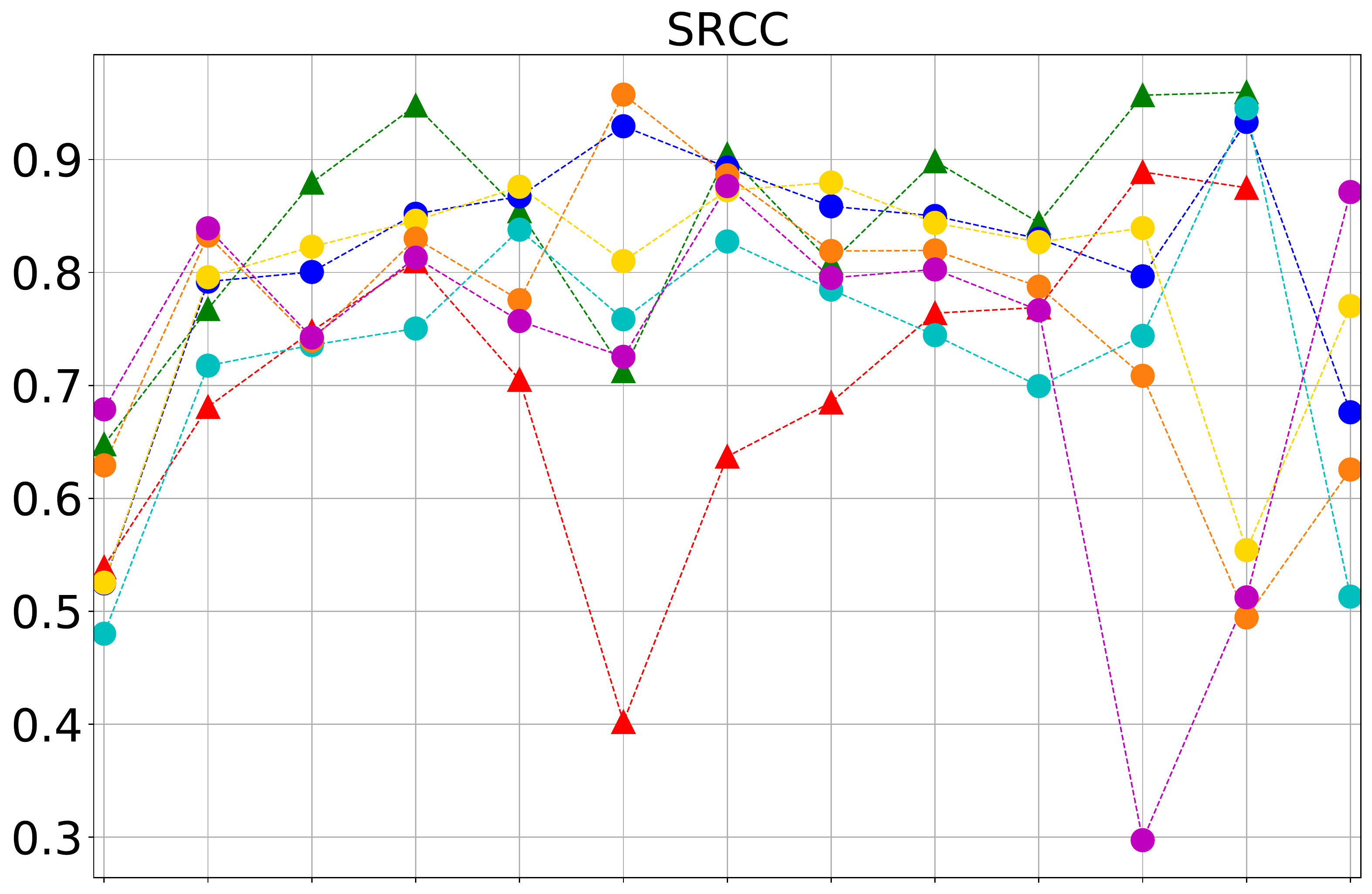}&
\includegraphics[width=0.65\columnwidth]{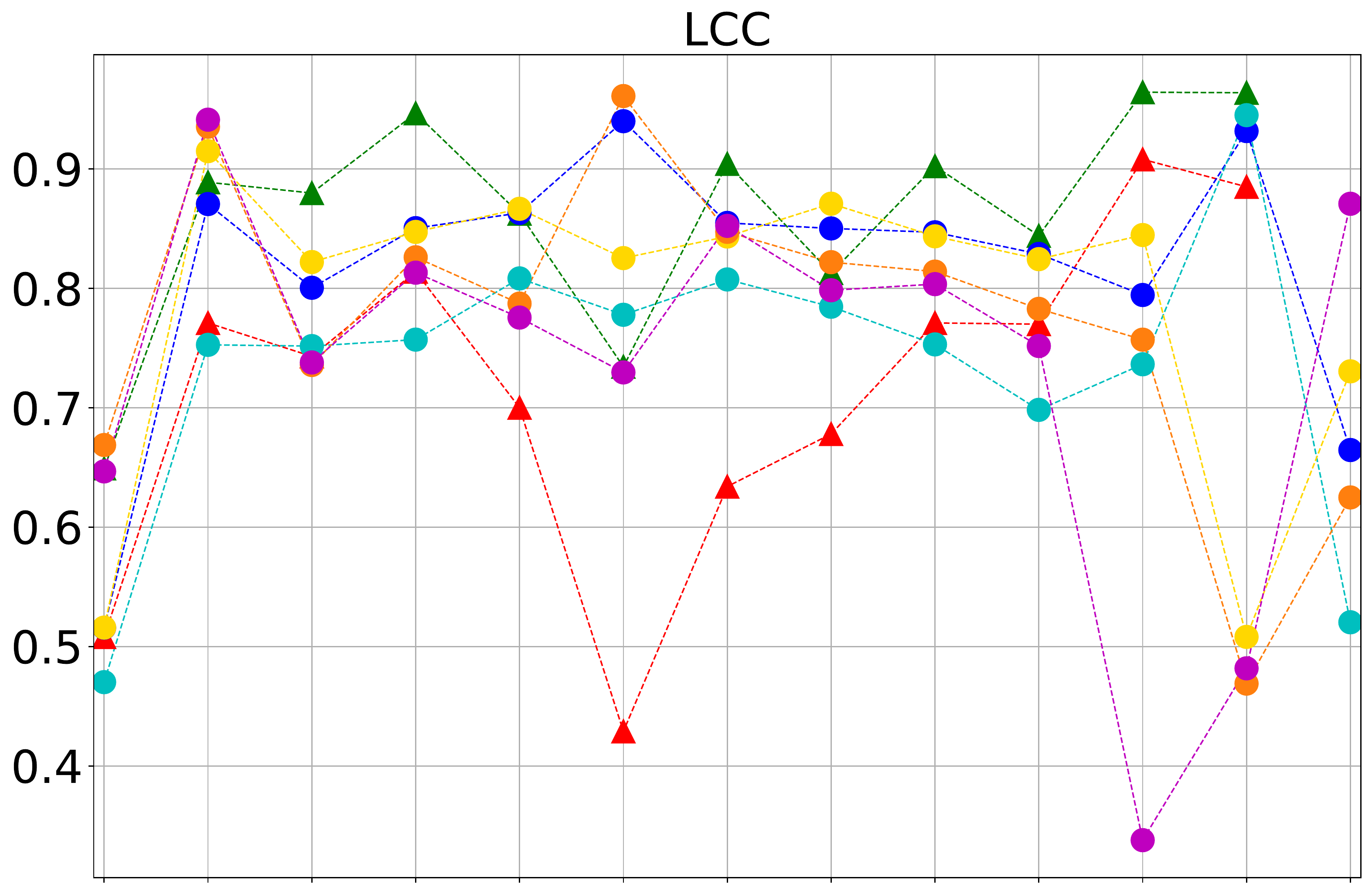}&
\includegraphics[width=0.65\columnwidth]{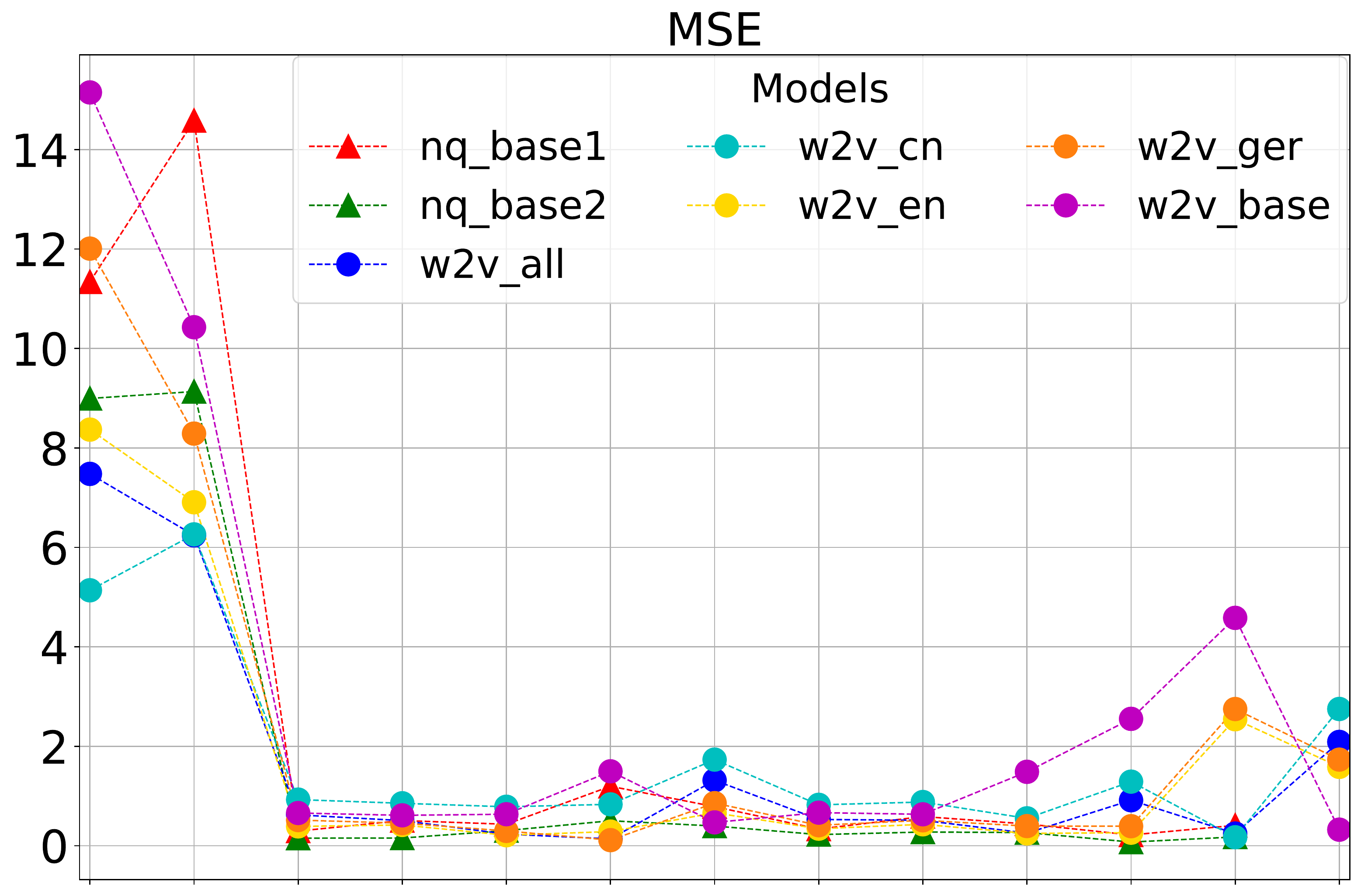}\\
\includegraphics[width=0.65\columnwidth]{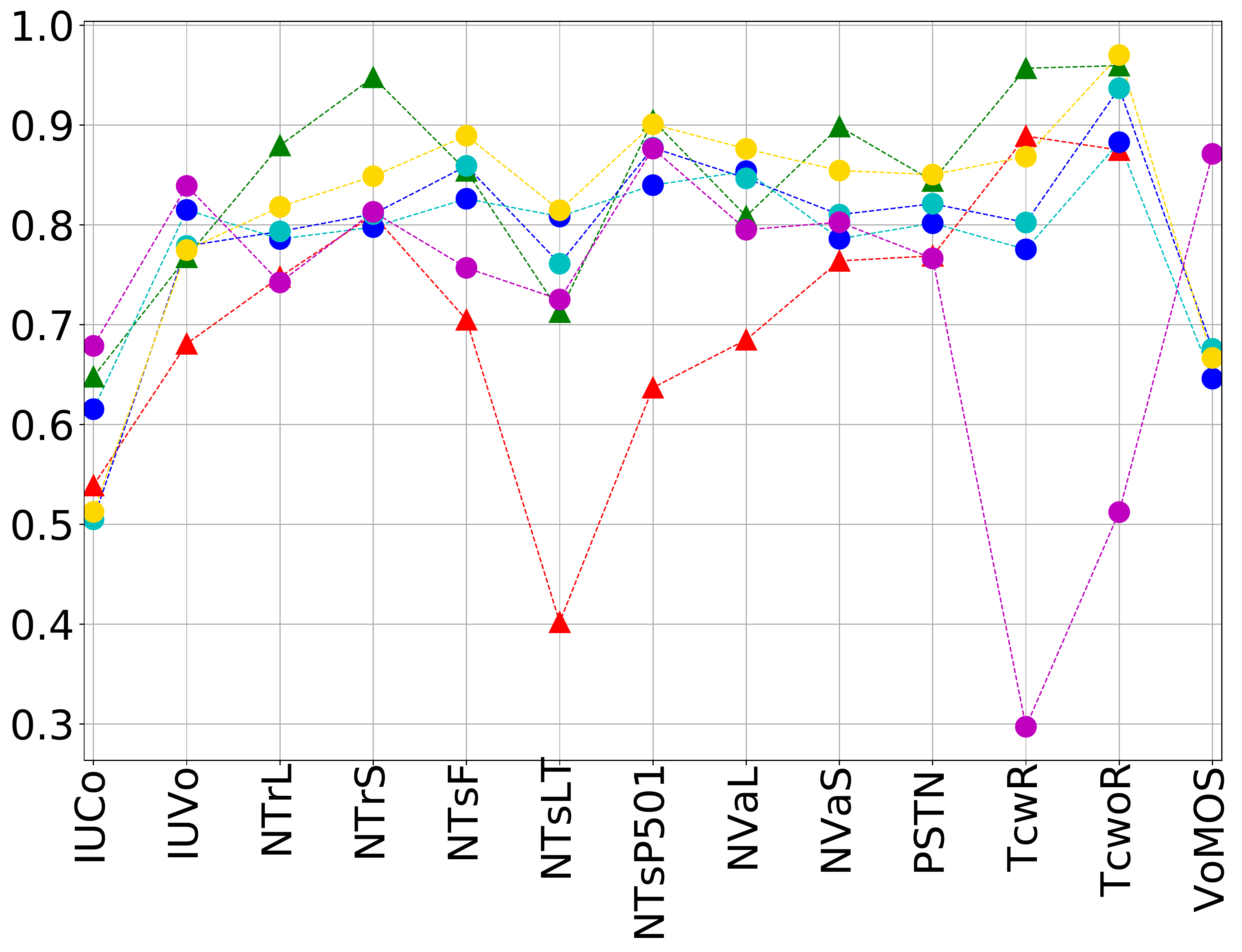}&
\includegraphics[width=0.65\columnwidth]{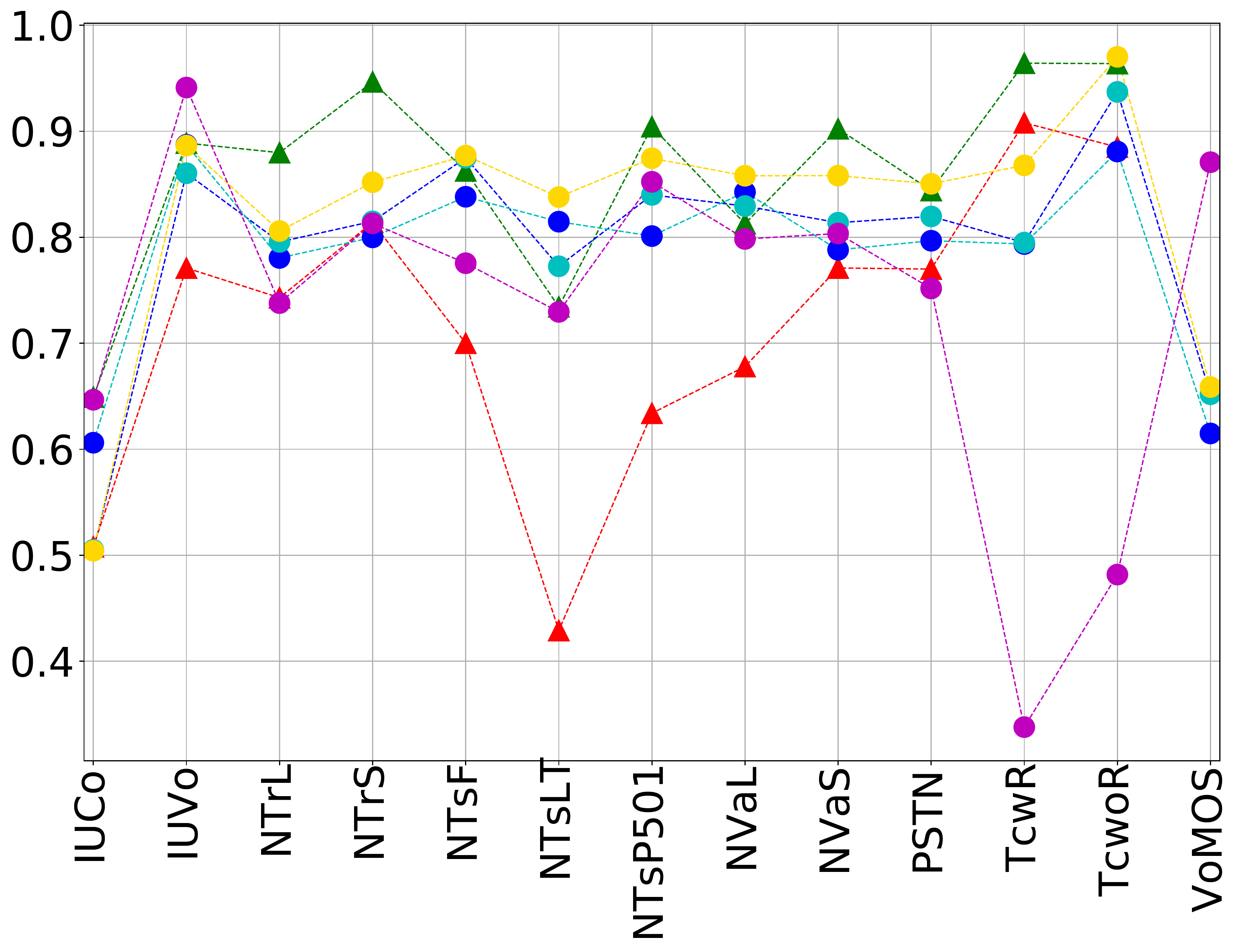}&
\includegraphics[width=0.65\columnwidth]{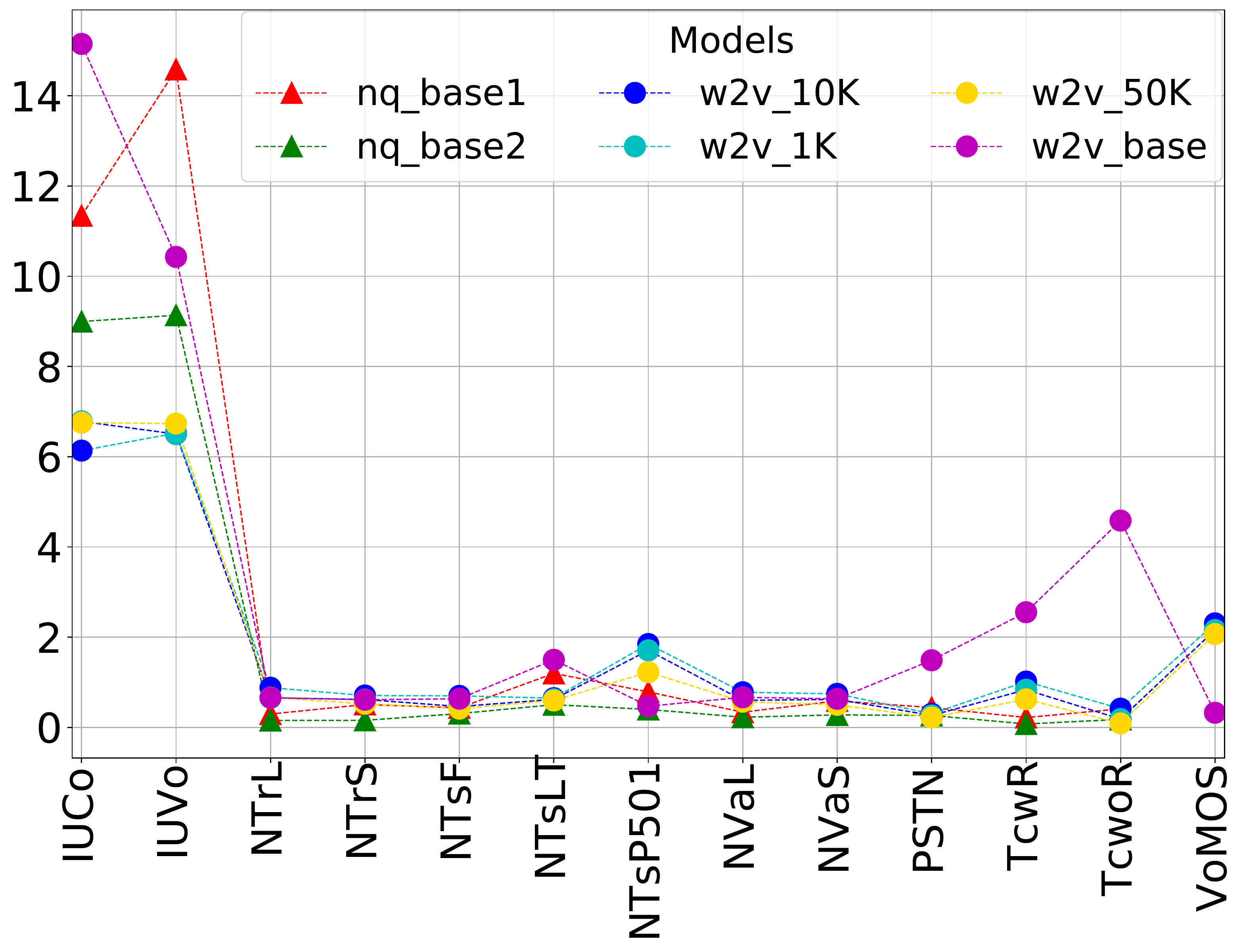}
\end{tabular}
\caption{SRCC, LCC and MSE results for all fine-tuned and baseline models across all speech datasets. First row: Experiment 1 - language influence. Second row: Experiment 2 - size influence.}\label{fig:results}
\end{figure*}

This section presents the results of the fine-tuned models across all 13 datasets included in this study. These models are compared against wav2vec 2.0 fine-tuned with the VoiceMOS dataset (w2v\_base) and two NISQA based models (nq\_base1 and nq\_base2) trained both over NISQA, PSTN and Tencent datasets with different architecture settings. The selection of w2v\_base as a baseline was led by its good performance over synthesised speech data reported in \cite{cooper2021generalization}. The performance of these models was evaluated using the MSE, the Spearman’s rank correlation coefficient (SRCC) and the linear correlation coefficient (LCC). Results for all fine-tuned and baseline models were grouped for the two experiments, and they are presented in Figure \ref{fig:results}. The first row presents the performance of the models for the language fine-tuning; meanwhile, the second row shows the results for the size fine-tuning aspect. Evaluated datasets are listed in the x~axis.

\subsection{Experiment 1: Language analysis}

Among the fine-tuned models generated for this experiment, w2v\_en and w2v\_all showed good performance across most NISQA and PSTN corpora (English speech samples). The w2v\_en model, fine-tuned exclusively with English samples, outperformed the other models in terms of SRCC and LCC at the NISQA ValLive dataset. Meanwhile, the w2v\_all model shows its best individual SRCC and LCC scores on the NISQA TestLiveTalk and the Tencent w/o Reverberation datasets, German and Chinese speech datasets, respectively. The w2v\_ger model surpassed all other models when tested against the NISQA TestLive corpus, a dataset containing only German speech. As for w2v\_cn, its best individual performance was reported for the Tencent w/o Reverberation corpus (Chinese speech samples). Despite the good performance observed for the fine-tuned models, baseline models still surpass them at several datasets. nq\_base2 reported the best correlation scores across several NISQA, PSTN and Tencent corpora. As for the w2v\_base model, its best performance was observed for the IU-Bloomington corpora and the VoiceMOS dataset; yet, it surprises its poor performance against the Tencent w Reverberation dataset, which was the worst correlation performance across all models. Finally, the nq\_base1 model showed a poor performance across most datasets, and it reported its worst individual correlation performance for the NISQA TestLiveTalk dataset.

From an MSE perspective, IU-Bloomington Cosine and Voices datasets showed the highest values across all models. This is explained in the methodology adopted for the subjective experiment (MUSHRA method) and its following re-scaling procedure. Aside from those datasets, most models reported MSE scores below 1 for most datasets. For both Tencent datasets, the w2v\_base model showed MSE values up to 4, which evidence the model's difficulty in dealing with non-English speech samples. This experiment showed that fine-tuning with different languages helped the model generalize better across other datasets. This is evidenced by the enormous gap between the w2v\_all and w2v\_base models, especially over non-English datasets.

\subsection{Experiment 2: Size analysis}

For the second experiment, three models were fine-tuned using different amounts of labelled data containing English and Chinese speech samples. Among these models, it is observed that the model w2v\_50K consistently reports better performance than the models fine-tuned with 10 k and 1 k samples. However, the performance gap between these three models is not very pronounced; they all show very similar correlation scores across all 13 datasets. The w2v\_50K model surpasses all other models on at least 3 NISQA datasets (TestFor, TestLiveTalk and ValLive), the Tencent w/o Reverberation and the PSTN dataset. Compared to the baseline models, nq\_base2 outperforms w2v\_50K at 3 NISQA corpora (TrainLive, TrainSim and ValSim) and the Tencent w Reverberation. These models showed a very narrow performance gap, except for the NISQA TrainLive and TrainSim and the Tencent w Reverberation datasets. As for the w2v\_base model, its best performance was reported for the IU-Bloomington datasets (Cosine and Voices) and the VoiceMOS corpus. As in the previous experiment, MSE performance reported values around 1 for all datasets and models, except for the IU-Bloomington datasets.

The results from this experiment showed that increasing the size of the subset to fine-tune a model will certainly benefit the model's performance, even at the point of surpassing strong baseline models at some datasets. However, it was also observed that fine-tuning, even with smaller datasets (1 k and 10 k), results in a relatively small cost in performance. This can be beneficial for a context in which labelled data is scarce, and a slight drop in prediction accuracy is accepted.

\subsection{Overall Analysis}
\begin{figure}[t]
\centering
\includegraphics[width=\columnwidth]{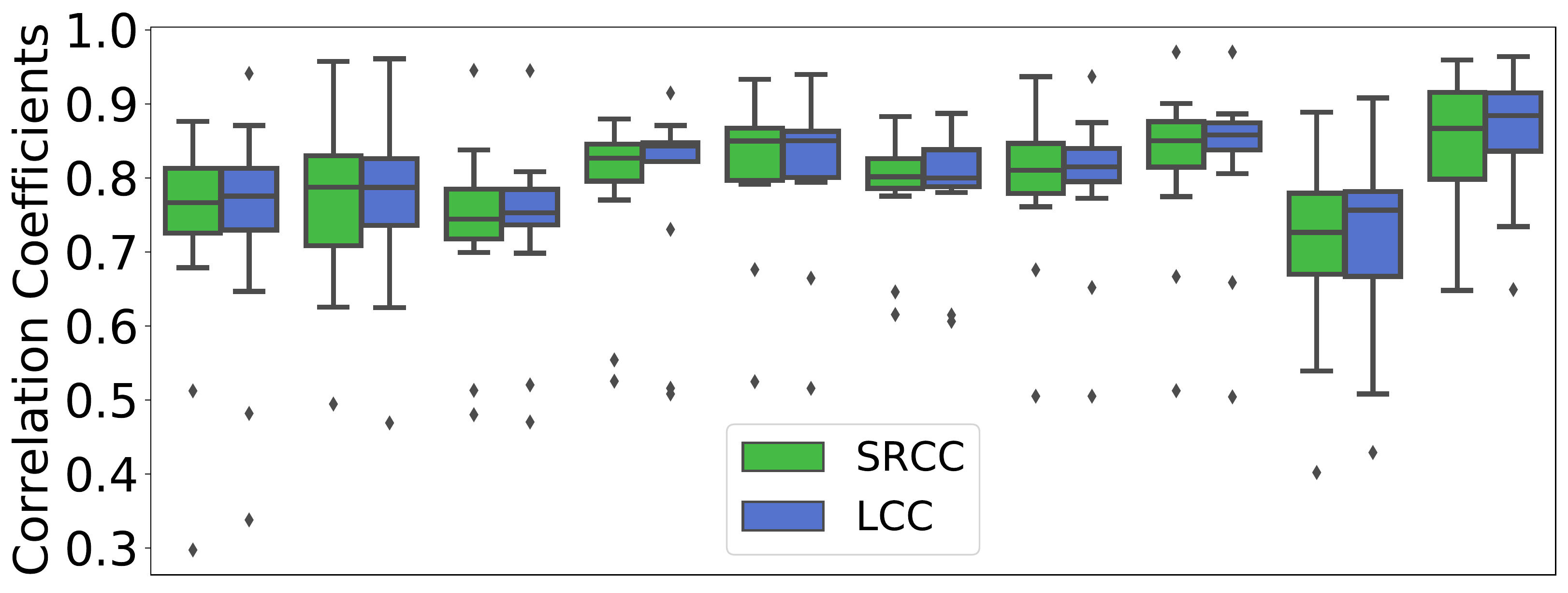}\\
\includegraphics[width=\columnwidth]{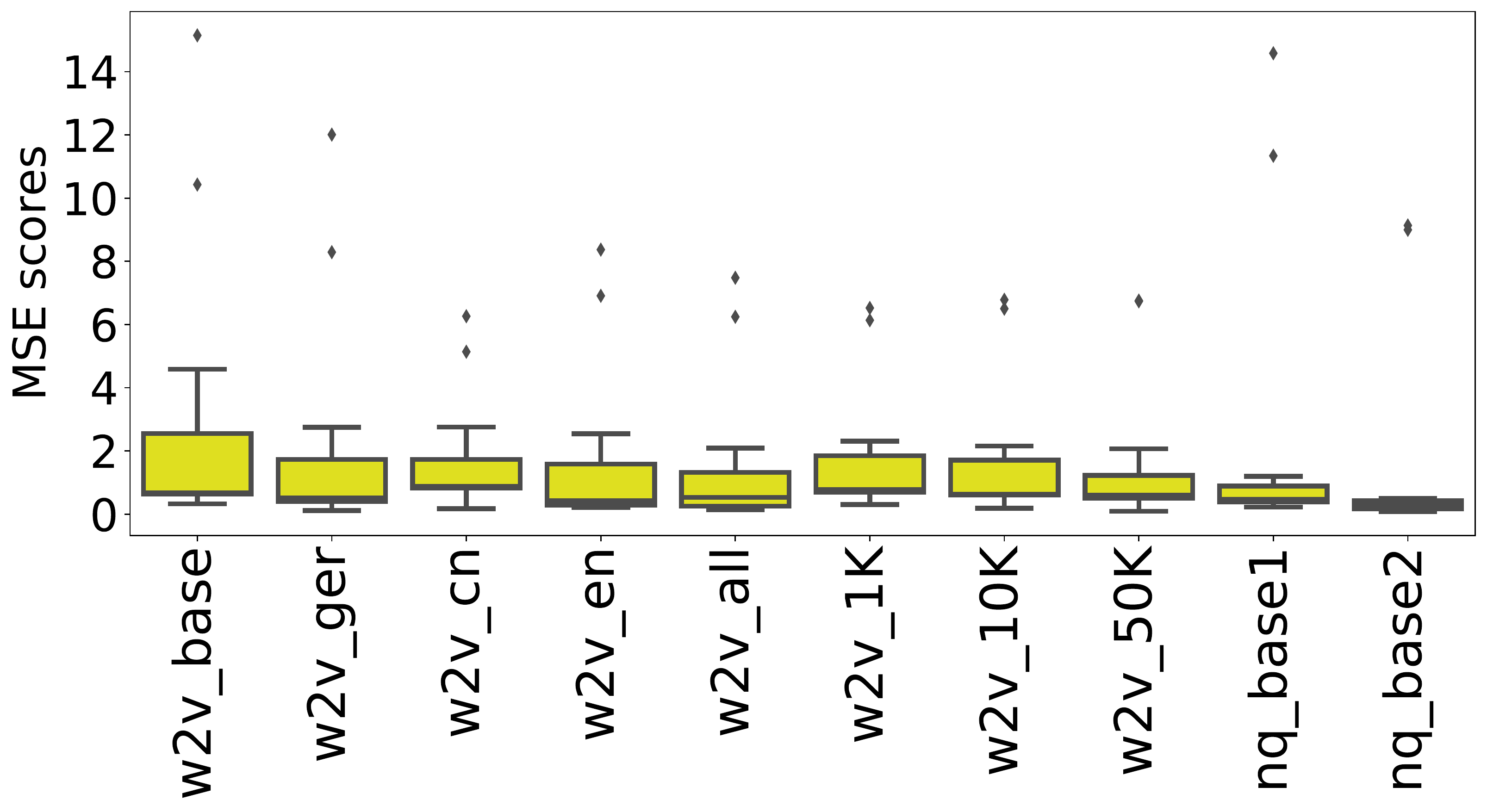}\\
\caption{Aggregated performance results across datasets.}
\label{fig:result_boxplot}
\end{figure}

In order to understand how wav2vec 2.0 fine-tuned models behave compared to NISQA based models, performance results from all 13 datasets were grouped and analysed in this section. Individual SRCC, LCC and MSE scores for each fine-tuned and baseline model across 13 datasets were averaged, and its distribution is depicted on the boxplots presented in Figure \ref{fig:result_boxplot}. The figure presents the distribution of the SRCC and LCC coefficients (first row) and the MSE scores (second row) across all 13 datasets for each evaluated model (x axis). It is observed that most of the fine-tuned models, except for the w2v\_ger model, report a low variance among their correlation scores. Fine-tuned models exhibit a more stable performance across different datasets despite being individually surpassed at several datasets. Compared to the baseline models, w2v\_all and w2v\_50K reported almost equivalent median values than the nq\_base2 baseline model but with lower variance across datasets. In terms of MSE, it is clear that NISQA based models produce closer predictions to the MOS ground truth data. Among the fine-tuned models, it is observed how the size of the dataset helps the model at making predictions with fewer errors (narrower variance towards the w2v\_50K model).

Regarding the effect of the degradation type on the models’ performance, no unexpected patterns were observed across datasets. However, it was observed that IU-Bloomington Cosine, a dataset containing samples degraded by background noise, received the lowest scores across all models. A similar condition was observed for VoiceMOS dataset, where most models, except for w2v\_base, reported low correlation scores and high error rates. Which was expected, since w2v\_base was fine-tuned using synthetic speech.

%Figure

\section{Discussion}

In this paper, two sets of models were generated by fine-tuning the wav2vec 2.0 pre-trained model targeting different languages and dataset sizes. These models were tested against one w2v\_base model and two NISQA baseline models across 13 different datasets. Results showed that nq\_base2 model performs best at most datasets; however, w2v\_all and w2v\_50K reported similar performance with lower variance across several datasets.

Findings from Experiment 1, where the language aspect of the fine-tuning step was evaluated, showed that performance from wav2vec 2.0 based models could be benefit from including speech samples from different languages, even in small amounts. These results align with the findings presented in \cite{cooper2021generalization}, where wav2vec 2.0 models were evaluated over datasets containing Chinese and Japanese speech samples. In their study, the xlsr model, a wav2vec 2.0 based model pre-trained with almost 56 k hours of speech containing 53 different languages \cite{conneau2020unsupervised}, showed good generalization ability. Multi-language pre-trained wav2vec 2.0 models can lead to the development of more robust models that could deal with language bias more efficiently \cite{Jimenez21language}. In \cite{conneau2020unsupervised}, a cluster analysis of the speech representations produced by the xlsr pre-train model showed that languages that share a similar root, like English and German or Italian and Spanish, tended to cluster together. This can benefit languages that have scarce annotated material as they can rely on other similar languages to fine-tune and obtain higher prediction performance. High-performance techniques that require less annotated data are highly needed, and they demand further exploration \cite{ragano2021more}. Moreover, studies targeting the pre-training step would reveal how accurate a model can get using a language-dedicated pre-trained model.

Experiment 2 aimed to explore how the dataset size used to fine-tune the models affected the overall performance. The findings from this experiment showed that, although a larger amount of samples will guarantee higher correlation and lower error rates, the performance gap is not necessarily large. This behaviour, where wav2vec 2.0 models report stable and moderate correlations, was also observed in \cite{cooper2021generalization}. In contrast, the performance of NISQA baseline models, trained with different architecture settings, showed noteworthy differences across the evaluated datasets. These baseline models show the best and worst performances for several datasets. A similar fine-tuning experiment using NISQA would allow us to explore and understand how is the size aspect affects NISQA.

\section{Conclusions}

This study explored the wav2vec 2.0 pre-trained model and the influence of fine-tuning it with different languages and data sizes. Based on the findings of this study, it is possible to conclude that fine-tuning a wav2vec 2.0 model using 1) a mid-size labelled dataset and 2) with a good language representation in the fine-tuning dataset will produce competitive models with steady performance. Better performance can be achieved by increasing the size of the fine-tune dataset with minor variance. Such models might be capable of competing against strong models trained with larger datasets like NISQA based models evaluated in this study. Similar wav2vec 2.0 models, pre-trained with multi-lingual datasets, need to be explored to build more resilient models across different languages.

Overall, this study showed the feasibility of tailoring wav2vec 2.0 for specific scenarios to perform well with limited annotated data.  Within Quality of Experience, a research area where labelled data is a critical limitation, wav2vec stands as a promising alternative to deal with setups involving audio or even explore its application for other media types.

\section{Acknowledgements}
This publication has emanated from research conducted with
the financial support of Science Foundation Ireland (SFI) under Grant Number 17/RC/2289\_P2 and 17/RC-PhD/3483. For the purpose of Open Access, the authors have applied a CC BY public copyright licence to any Author Accepted Manuscript version arising from this submission.

\bibliographystyle{IEEEtran}

\bibliography{mybib}

% Generated by IEEEtran.bst, version: 1.13 (2008/09/30)
\begin{thebibliography}{10}
\providecommand{\url}[1]{#1}
\csname url@samestyle\endcsname
\providecommand{\newblock}{\relax}
\providecommand{\bibinfo}[2]{#2}
\providecommand{\BIBentrySTDinterwordspacing}{\spaceskip=0pt\relax}
\providecommand{\BIBentryALTinterwordstretchfactor}{4}
\providecommand{\BIBentryALTinterwordspacing}{\spaceskip=\fontdimen2\font plus
\BIBentryALTinterwordstretchfactor\fontdimen3\font minus
  \fontdimen4\font\relax}
\providecommand{\BIBforeignlanguage}[2]{{%
\expandafter\ifx\csname l@#1\endcsname\relax
\typeout{** WARNING: IEEEtran.bst: No hyphenation pattern has been}%
\typeout{** loaded for the language `#1'. Using the pattern for}%
\typeout{** the default language instead.}%
\else
\language=\csname l@#1\endcsname
\fi
#2}}
\providecommand{\BIBdecl}{\relax}
\BIBdecl

\bibitem{rec1996p}
I.~Rec, ``P. 800: Methods for subjective determination of transmission
  quality,'' \emph{International Telecommunication Union, Geneva}, vol.~22,
  1996.

\bibitem{recommendation2001method}
I.~Recommendation, ``Method for the subjective assessment of intermediate sound
  quality (mushra),'' \emph{ITU, BS}, pp. 1543--1, 2001.

\bibitem{hines2012speech}
A.~Hines and N.~Harte, ``Speech intelligibility prediction using a neurogram
  similarity index measure,'' \emph{Speech Communication}, vol.~54, no.~2, pp.
  306--320, 2012.

\bibitem{beerends2013perceptual}
J.~G. Beerends, C.~Schmidmer, J.~Berger, M.~Obermann, R.~Ullmann, J.~Pomy, and
  M.~Keyhl, ``Perceptual objective listening quality assessment (polqa), the
  third generation itu-t standard for end-to-end speech quality measurement
  part i—temporal alignment,'' \emph{Journal of the Audio Engineering
  Society}, vol.~61, no.~6, pp. 366--384, 2013.

\bibitem{malfait2006p}
L.~Malfait, J.~Berger, and M.~Kastner, ``P. 563—the itu-t standard for
  single-ended speech quality assessment,'' \emph{IEEE Transactions on Audio,
  Speech, and Language Processing}, vol.~14, no.~6, pp. 1924--1934, 2006.

\bibitem{tseng2021utilizing}
W.-C. Tseng, C.~yu~Huang, W.-T. Kao, Y.~Y. Lin, and H.~yi~Lee, ``{Utilizing
  Self-Supervised Representations for MOS Prediction},'' in \emph{Proc.
  Interspeech 2021}, 2021, pp. 2781--2785.

\bibitem{cooper2021generalization}
E.~Cooper, W.-C. Huang, T.~Toda, and J.~Yamagishi, ``Generalization ability of
  mos prediction networks,'' \emph{arXiv preprint arXiv:2110.02635}, 2021.

\bibitem{mittag2021nisqa}
G.~Mittag, B.~Naderi, A.~Chehadi, and S.~Möller, ``{NISQA: A Deep
  CNN-Self-Attention Model for Multidimensional Speech Quality Prediction with
  Crowdsourced Datasets},'' in \emph{Proc. Interspeech 2021}, 2021, pp.
  2127--2131.

\bibitem{jimenez2021removing}
R.~Z. Jim{\'e}nez, G.~Mittag, and S.~M{\"o}ller, ``Removing the bias in speech
  quality scores collected in noisy crowdsourcing environments,'' in \emph{2021
  13th International Conference on Quality of Multimedia Experience
  (QoMEX)}.\hskip 1em plus 0.5em minus 0.4em\relax IEEE, 2021, pp. 49--54.

\bibitem{Jimenez21language}
\BIBentryALTinterwordspacing
R.~Z. Jim{\'{e}}nez, B.~Naderi, and S.~M{\"{o}}ller, ``Influence of language
  differences in crowdsourcing speech quality assessment studies,'' in
  \emph{13th International Conference on Quality of Multimedia Experience,
  QoMEX 2021, Montreal, Canada, June 14-17, 2021}.\hskip 1em plus 0.5em minus
  0.4em\relax {IEEE}, 2021, pp. 43--48. [Online]. Available:
  \url{https://doi.org/10.1109/QoMEX51781.2021.9465437}
\BIBentrySTDinterwordspacing

\bibitem{chinen21marginal}
M.~Chinen, ``Marginal effects of language and individual raters on speech
  quality models,'' \emph{IEEE Access}, vol.~9, pp. 127\,320--127\,334, 2021.

\bibitem{fan2020exploring}
Z.~Fan, M.~Li, S.~Zhou, and B.~Xu, ``{Exploring wav2vec 2.0 on Speaker
  Verification and Language Identification},'' in \emph{Proc. Interspeech
  2021}, 2021, pp. 1509--1513.

\bibitem{stupakov2009cosine}
A.~Stupakov, E.~Hanusa, J.~Bilmes, and D.~Fox, ``Cosine-a corpus of multi-party
  conversational speech in noisy environments,'' in \emph{2009 IEEE
  International Conference on Acoustics, Speech and Signal Processing}.\hskip
  1em plus 0.5em minus 0.4em\relax IEEE, 2009, pp. 4153--4156.

\bibitem{richey2018voices}
C.~Richey, M.~A. Barrios, Z.~Armstrong, C.~Bartels, H.~Franco, M.~Graciarena,
  A.~Lawson, M.~K. Nandwana, A.~Stauffer, J.~{van Hout}, P.~Gamble,
  J.~Hetherly, C.~Stephenson, and K.~Ni, ``{Voices Obscured in Complex
  Environmental Settings (VOiCES) Corpus},'' in \emph{Proc. Interspeech 2018},
  2018, pp. 1566--1570.

\bibitem{mittag2020dnn}
G.~Mittag, R.~Cutler, Y.~Hosseinkashi, M.~Revow, S.~Srinivasan, N.~Chande, and
  R.~Aichner, ``{DNN No-Reference PSTN Speech Quality Prediction},'' in
  \emph{Proc. Interspeech 2020}, 2020, pp. 2867--2871.

\bibitem{baevski2020wav2vec}
A.~Baevski, Y.~Zhou, A.~Mohamed, and M.~Auli, ``wav2vec 2.0: A framework for
  self-supervised learning of speech representations,'' \emph{Advances in
  Neural Information Processing Systems}, vol.~33, pp. 12\,449--12\,460, 2020.

\bibitem{baevski2020effectiveness}
A.~Baevski and A.~Mohamed, ``Effectiveness of self-supervised pre-training for
  asr,'' in \emph{ICASSP 2020-2020 IEEE International Conference on Acoustics,
  Speech and Signal Processing (ICASSP)}.\hskip 1em plus 0.5em minus
  0.4em\relax IEEE, 2020, pp. 7694--7698.

\bibitem{panayotov2015librispeech}
V.~Panayotov, G.~Chen, D.~Povey, and S.~Khudanpur, ``Librispeech: an asr corpus
  based on public domain audio books,'' in \emph{2015 IEEE international
  conference on acoustics, speech and signal processing (ICASSP)}.\hskip 1em
  plus 0.5em minus 0.4em\relax IEEE, 2015, pp. 5206--5210.

\bibitem{conneau2020unsupervised}
A.~Conneau, A.~Baevski, R.~Collobert, A.~Mohamed, and M.~Auli, ``{Unsupervised
  Cross-Lingual Representation Learning for Speech Recognition},'' in
  \emph{Proc. Interspeech 2021}, 2021, pp. 2426--2430.

\bibitem{ragano2021more}
A.~Ragano, E.~Benetos, and A.~Hines, ``More for less: Non-intrusive speech
  quality assessment with limited annotations,'' in \emph{2021 13th
  International Conference on Quality of Multimedia Experience (QoMEX)}.\hskip
  1em plus 0.5em minus 0.4em\relax IEEE, 2021, pp. 103--108.

\end{thebibliography}

% \begin{thebibliography}{9}
% \bibitem[1]{Davis80-COP}
%   S.\ B.\ Davis and P.\ Mermelstein,
%   ``Comparison of parametric representation for monosyllabic word recognition in continuously spoken sentences,''
%   \textit{IEEE Transactions on Acoustics, Speech and Signal Processing}, vol.~28, no.~4, pp.~357--366, 1980.
% \bibitem[2]{Rabiner89-ATO}
%   L.\ R.\ Rabiner,
%   ``A tutorial on hidden Markov models and selected applications in speech recognition,''
%   \textit{Proceedings of the IEEE}, vol.~77, no.~2, pp.~257-286, 1989.
% \bibitem[3]{Hastie09-TEO}
%   T.\ Hastie, R.\ Tibshirani, and J.\ Friedman,
%   \textit{The Elements of Statistical Learning -- Data Mining, Inference, and Prediction}.
%   New York: Springer, 2009.
% \bibitem[4]{YourName17-XXX}
%   F.\ Lastname1, F.\ Lastname2, and F.\ Lastname3,
%   ``Title of your INTERSPEECH 2022 publication,''
%   in \textit{Interspeech 2022 -- 23\textsuperscript{rd} Annual Conference of the International Speech Communication Association, September 18-22, Incheon, Korea, Proceedings, Proceedings}, 2022, pp.~100--104.
% \end{thebibliography}

\end{document}